\documentclass[prl,aps,twocolumn,groupedaddress,showpacs,floatfix,preprintnumbers,superscriptaddress]{revtex4-1}
\usepackage{amsmath,amssymb,multirow,bm}
\usepackage[colorlinks,linkcolor=blue,anchorcolor=blue,citecolor=blue,urlcolor=blue,filecolor=black]{hyperref}

\newcommand{\beq}{\begin{equation}}
\newcommand{\eeq}{\end{equation}}
\newcommand{\bea}{\begin{eqnarray}}
\newcommand{\eea}{\end{eqnarray}}
\usepackage{graphicx}

\begin{document}


\title{Nuclear Spectroscopy with Heavy Ion Nucleon Knockout and (p,2p) Reactions}

\author{Jianguo Li}
\email[]{Email: jianguo\_li@pku.edu.cn}
\affiliation{School of Physics, and State Key Laboratory of Nuclear Physics and Technology, Peking University, Beijing 100871, China}

\author{Carlos A. Bertulani}
\email[]{Email: carlos.bertulani@tamuc.edu}
\affiliation{Department of Physics and Astronomy, Texas A\&M University-Commerce,  TX 75429-3011, USA}
\affiliation{Institut f\"ur Kernphysik, Technische Universit\"at Darmstadt, D-64289 Darmstadt, Germany}

\author{Furong Xu}
\email[]{Email: frxu@pku.edu.cn}
\affiliation{School of Physics, and State Key Laboratory of Nuclear Physics and Technology, Peking University, Beijing 100871, China}

\begin{abstract}
Knockout reactions with heavy ion targets in inverse kinematics, as well as ``quasi-free" (p,2p) and (p,pn) reactions are useful tools for nuclear spectroscopy.  We report calculations on \textit{ab-initio} many-body wavefunctions based on the no-core shell model  to study the nucleon removal reactions in light nuclei, including beryllium, carbon, and oxygen isotopic chains, and explore the importance of using an \textit{ab-initio} method.   Our study helps clarifying how the extraction of spectroscopic factors from the experiments depend on the details of the many-body wavefunctions being probed. We show that recent advances with the ab-initio method can provide more insights on the spectroscopy information extracted from experiments.
\end{abstract}

\maketitle

\section{Introduction}

Heavy ion nucleon reactions in which an impinging nucleus has one of its nucleons removed by a hard collision with a target nucleus, have become a standard spectroscopic tool. The so-called nucleon knockout is particularly useful to study reactions with radioactive nuclear beams. Another much celebrated spectroscopic tool is (p,2p) or (p,pn) reactions, using hydrogen targets for studies involving nuclei far from the stability \cite{AUMANN2021103847}. Various subjects of interest for nuclear physics have been assessed with knockout and quasi-free, (p,2p) and (p,pn), reactions such as magicity, shell evolution, the structure of loosely-bound nuclei, short-range correlations \cite{PhysRevLett.122.172502,Patsuyk21}, etc. Since the first experimental campaigns using radioactive beams in knockout reactions \cite{RIISAGER1992365,Orr:92,Bazin:2009}, the community has used absolute value of cross sections, as well as the momentum distributions of the fragment core to identify the quantum numbers of the removed nucleon, as well as the details of the nuclear  wavefunctions  \cite{Flavigny:2012,doi:10.1063/1.4909557,PhysRevC.103.054610,PhysRevC.103.054610,Tsang09,otsuka2020,lee2011,pty011,PhysRevC.100.044609,PhysRevC.93.034333,PhysRevLett.104.112701,Atar:2018,Panin:16,DiazPRC.97.024311,PANIN2019134802,HOLL2019682}.

In direct reactions, the amplitude of the overlap function of the bound-state wave functions of the initial and final nuclei is the telltale of the total nuclear wave functions and interaction potentials. The overlap functions are defined as
\beq
\begin{aligned}
I_{lj}(r) = & \langle \Psi_A^{J_A}| \left [|\Psi_{A-1}^{J_{A-1}}\rangle \otimes |l,j\rangle^{J_A} \right]\rangle, \\
          = &  \sum_i \langle \Psi_A^{J_A}||a_{n_ilj}^{\dagger}|| \Psi_{A-1}^{J_{A-1}}\rangle \langle n_ilj|u_i\rangle,
\end{aligned}
\label{overlap}
\eeq
where $|\Psi_A^{J_A}\rangle$ and $|\Psi_{A-1}^{J_{A-1}}\rangle$ are wave functions of the nuclei $A$ and $A-1$, respectively.  The $a_{n_ilj}^{\dagger}$ is a creation operator associated with the single particle basis state $|u_i\rangle$.
Within Eq. (\ref{overlap}), the overlap function involves summation over all the single-particle states with the same $lj$ quantum number. The final integrals have small dependence of the single-particle basis assumed, which are in contrast to the standard shell-model  calculations within a limited model space where the model-dependence enters through
the specific choice of a single particle state $|nlj\rangle$, with Eq. \ref{overlap} reducing to a single matrix elements $\langle \Psi_A^{J_A}||a_{nlj}^{\dagger}(i)|| \Psi_{A-1}^{J_{A-1}}\rangle$ \cite{RevModPhys.77.427}.
Note that $I_{lj}(r)$ is  not an eigenfunction of a Hermitian Hamiltonian and cannot be directly associated with a probability. Thus, it is not normalized to unity and the spectroscopic factor for the nuclear configuration, defined as
\begin{equation}
C^2S_{lj} = \int dr \, r^2 |I_{lj}(r)|^2,
\label{specfac}
\end{equation}
can be larger than unity. The $C^2S$  is a model-dependent quantity that can be calculated in the shell-model, being sensitive to the interactions and to the truncations of the model space. They usually differ from unity because they depends on the contribution of numerous antisymmetrized and different nonorthogonal channels coupled to the two-body $n(A-1)$ channel.

At a large distance $r$, the overlap function is proportional to the Whittaker function, depending only on the charge of the particles, the binding energy, and a normalization constant, i.e.,
\begin{equation}
I_{lj}(r) \longrightarrow C_{lj} {1\over r} W_{-\eta,l+1/2} (2\kappa r),
\end{equation}
with $\mu$ the nucleon+$(A-1)$ nucleus reduced mass, $\eta=\mu Z_n Z_{A-1} e^2/\hbar \kappa$ the Sommerfeld parameter, $\kappa=\sqrt{2\mu E_B}/\hbar$  the wavenumber,  $E_B$ the nucleon separation energy,  $Z_{A-1}$ and $Z_n$ the residual nucleus and nucleon charges, and $l$ the nucleon angular momentum. Thus, at large distances, $|I_{lj}|^2$ should be proportional to the square of the normalization coefficients (ANC): $C_{lj}^2$. It is often stated that, due to their peripheral character, heavy ion knockout reactions are directly proportional to $C^2$, although this view is not confirmed by detailed calculations \cite{BertulaniIdini}. In fact, it was already noted that the single-neutron knockout reactions from  $^{10}$C and $^{10}$Be beams incident on light nuclear targets have demonstrated a sensitivity to differences in shell-model, no-core shell model (NCSM), and Variational Monte Carlo (VMC) wavefunctions and therefore may help the development of ab initio structure models \cite{PhysRevLett.106.162502}. In this work we will concentrate on how theory can provide an accurate account of the spectroscopic quantities, defined in Eqs. \eqref{overlap} and \eqref{specfac}. This is timely, as a ``quenching" of the spectroscopic factors in  heavy ion knockout reactions has become a hot topic in the recent literature \cite{PhysRevC.90.057602,PhysRevC.103.054610,AUMANN2021103847}, although (p,2p) reactions seem to contradict these findings \cite{Pan16,Atar:2018,DiazPRC.97.024311,HOLL2019682}. Here we explore the impact on the experimental analysis in the case one uses ab initio wavefunctions to calculate both heavy ion knockout and (p,2p) reactions.

New experiments using $(p,pN)$  reactions, with $N = p, n$, in inverse kinematics have been reported \cite{Pan16,Atar:2018,DiazPRC.97.024311,HOLL2019682}, being accompanied by new developments in reaction theories using different models than those adopted for heavy ion knockout reactions. In particular, (p,pN) reactions are more sensitive to the inner part of the nuclear wavefunction \cite{BertulaniIdini}, specially for light nuclear projectiles  \cite{Aumann:13,Panin:16,Atar:2018}. This has been clearly discussed in Refs. \cite{Aumann:13,BertulaniIdini}. One thus expects that $(p,pN)$ reactions involve an increased sensitivity to the many-body aspects of the single particle content of the nuclear wavefunctions. The sole knowledge of spectroscopic factors is not enough for a good description of $(p,pN)$ reactions and in some cases even for heavy ion knockout reactions.
We use the reaction theory reported in Refs. \cite{Hansen:03,Bertulani:06} for the heavy ion knockout, and in Ref. \cite{Aumann:13} for the (p,2p) case. Other variants of these reactions models are found in Refs. \cite{Bertulani:92,PhysRevC.51.2646,Cravo:16,PhysRevC.53.2007,PhysRevC.54.3043,GOMEZRAMOS2018511,Bertulani:04,Ogata:15,LEI2021136032,Moro:15,PhysRevC.96.024609,PhysRevC.99.054622,Hebborn_2020,PhysRevC.104.024616}. The input of these calculations are the nucleon-nucleon cross sections, using the parametrization provided in Ref. \cite{BertulaniConti10}, and nuclear densities calculated with the ab initio procedure. It is worthwhile mentioning that the cross section obtained here depend very little ($\lesssim5\%$) on the differences between densities obtained with the ab initio and other methods, such as neutron/proton ratio scaled densities obtained from electron scattering experiments. Therefore, the calculations are a major probe of the nuclear wavefunctions and their respective overlap integrals, as in Eq. \eqref{overlap}.

\section{Overlap functions with an ab initio method}
\subsection{ \textit{Ab initio} wavefunctions}
The \textit{ab initio} NCSM \cite{BARRETT2013131} has been employed.
Contrary to the standard shell-model in which the calculations are performed within limited model space using an inner frozen core,
the NCSM calculations are done without a core and the model space is as large as possible to obtain the converged results within the ability of  super-computer \cite{BARRETT2013131,PhysRevC.100.054313,PhysRevC.104.024319}. The correct treatment of internucleon correlations is one of the strongest features of the NCSM calculations within a large model space. In contrast, only configuration mixings within a small model space are considered in the standard shell-model calculations.  The nucleon orbital occupations are spread over a large space in  the NCSM,  but in the standard shell-model only single-particle orbitals within the restricted model space are occupied. In the NCSM, the center-of-mass correction is made using Lawson method \cite{BARRETT2013131}.
In the present work, we perform the NCSM using the Daejeon16  interaction \cite{SHIROKOV201687} which could provide good descriptions of light nuclei,
to compute the energies, overlap functions and $C^2S$ values for the states. The results for these quantities are shown in Table. \ref{GS}.

\begin{widetext}
\begin{center}
\begin{table}[h]
\caption{Theoretical and experimental total energies ($E$, in MeV) and calculated $C^2S$ values for the nuclear states studied in this work. $N_{\text{max}}$ indicates the harmonic oscillator basis space used in the NCSM calculations. The nucleus (left) added by one proton or neutron becomes the nucleus on the right. The asterisk $*$ indicates that this coupling cannot exist. An oscillator basis
with $\hbar w = 15$ MeV is used for the Daejeon16 interaction.}
\setlength{\tabcolsep}{1.0mm}{
\begin{tabular}{lccccccccccc}
\hline \hline
nucleus & states  & $E_{\text{exp}}$    & $E_{\text{th}}$ &  $N_{\text{max}}$ & nucleus   & states & $E_{\text{exp}}$    & $E_{\text{th}}$ &  $N_{\text{max}}$  &  $C^2S(p_{3/2})$& $C^2S(p_{1/2})$ \\
\hline
$^6$He & $0^+$ & $-$29.27 & $-$28.98 & 12 & $^7$Li & $3/2^-$ & $-$39.25 & $-$39.08 & 10 &0.50 & * \\
$^6$Li & $1^+$ & $-$31.99 & $-$31.57 & 12 &        &      &        &    &    & 0.47 & 0.18\\
       & $0^+$ & $-$28.43 & $-$28.00 & 12 &    &   &  &  &  & 0.25 & *\\
$^7$Be & $3/2^-$ & $-$37.60 & $-$40.14 & 10 & $^8$B &$2^+$  & $-$37.74 & $-$39.99 & 10 &0.76 & 0.07 \\
       &$1/2^-$ & $-$37.17 & $-$39.89 &  10 &   &   &   &   &       & 0.23 & * \\
$^8$B &  $2^+$ & $-$37.74 & $-$39.99 & 10 &$^9$C &$3/2^-$ & $-$39.04 & $-$39.71 & 8 &0.88 & 0.04 \\
$^8$Li & $2^+$ & $-$41.28 & $-$42.85 & 10 &$^9$Li & $3/2^-$ & $-$45.34 & $-$45.79 & 8 & 0.90 & 0.04\\
       & $1^+$ & $-$40.30 & $-$41.89 &  10& &      &     &          &          &0.30 & 0.00 \\
$^9$Li & $3/2^-$ & $-$45.34 & $-$45.79 &  8 & $^{10}$Be &$0^+$ & $-$64.98 & $-$67.40 & 8  &1.38 & * \\
       & $1/2^-$ & $-$42.69 & $-$44.62 &  8 &           &    &          &          & & * & 0.42 \\
$^9$C  & $3/2^-$ & $-$39.04 & $-$39.71 & 8 & $^{10}$C & $0^+$ & $-$60.32 & $-$62.89 & 8 &1.34 & *\\
$^{11}$B & $3/2^-$  & $-$76.21 & $-$74.50 & 6 &$^{12}$C & $0^+$  & $-$92.16 & $-$91.30  & 6 & 2.69 & * \\
$^{11}$C & $3/2^-$  & $-$73.44 & $-$71.80 & 6&           &  &          &           & & 2.50 &  *\\
$^{13}$N & $1/2^-$ & $-$94.11 & $-$95.26 &   6 &$^{14}$O & $0^+$  & $-$98.73 & $-$99.05 & 6 & * & 1.53 \\
$^{13}$O & $3/2^-$ & $-$75.55 & $-$73.20 &  6 &         &   &          &         & & 
3.234&  *\\
$^{15}$N & $1/2^-$ & $-$115.49 & $-$115.43 &  6 & $^{16}$O & $0^+$  & $-$127.62 & $-$129.79 & 6 &  *& 1.58 \\
         & $3/2^-$ & $-$105.17 & $-$105.96 &  6&          &  &            &          &  &3.07 & * \\
$^{15}$O & $1/2^-$ & $-$111.96 & $-$113.22 & 6 &          &  & & & & * & 1.70 \\
        & $3/2^-$ & $-$105.78 & $-$103.30 &  6 &          &  &  & & & 3.21 & *\\
\hline \hline
\end{tabular}}\label{GS}
\end{table}
\end{center}
\end{widetext}

\begin{widetext}
\begin{center}
\begin{table}[!htb]
\caption{Proton(neutron) knockout with $^7$Li projectiles. The computed reduction factor, R, is  the ratio between the experimental and theoretical inclusive one-nucleon-removal cross sections. $C^2S$(th) indicates that the spectroscopic factors are computed using the NCSM. $C^2S$(VMC) denotes that the results are calculated with the VMC.}
\setlength{\tabcolsep}{2.06mm}{
\begin{tabular}{lccccccccc} \hline \hline
 Reaction          & $E_{beam}$  & $J^\pi$  &$j$& $S_p$[$S_n$]&  $C^2S$(th)& $C^2S$(VMC) &  $\sigma_{th}$&  $\sigma_{exp}$ & R  \\  \hline
    & MeV/nucleon & && MeV& & &mb&mb\\  \hline
$^{9}$Be($^{7}$Li,$^{6}$He) \cite{PhysRevC.86.024315} & 80 & $0^+$ &3/2&9.98 &0.496& 0.439 & 28.13&13.4 (7) & 0.476 (24)\\
 $^{9}$Be($^{7}$Li,$^{6}$Li) \cite{PhysRevC.86.024315} & 120 & $1^+$ &1/2&7.25&0.176 & 0.24 &15.21&  \\
 &  &  &3/2&&0.473& 0.47 & 28.91& \\
 & & $0^+$ &3/2&10.8&0.250 & 0.219 & 17.09& \\
  & &  inclusive & & && & 61.21&30.7 (18) & 0.501(29) \\
\hline \hline
\end{tabular}}\label{SF1}
\end{table}
\end{center}
\end{widetext}

It is well known that ab initio wavefunctions obtained from expansions in harmonic oscillator wavefunctions have a hard time in reproducing the large distance behavior of the nuclear states and a large number of basis functions need to be used for the purpose. This does not diminish the merit of using such wavefunctions neither for quasi-free $(p,pN)$ reactions nor for knockout reactions with heavy ions. The bulk (but not all) of the cross sections in heavy ion  knockout reactions are due to the tail of the overlap integrals. Therefore, this tail has to be reproduced well. This can be easily fixed \cite{Navratil:06plb,Navratil:06prc} by using a procedure that replaces ab initio wavefunctions at their tails by those with appropriate asymptotic behavior such as solutions of a WS model. A fit extending to the internal part of the ab initio overlap functions yield and adequate renormalization yield correct knockout cross sections. In the case of (p,pN) reactions, an accurate description of the overlap integral tail is not of relevance  for the total cross sections, as the bulk of the cross sections are due to the inner part of the overlap integral.

Following Ref. \cite{BrownPRC.65.061601}, we define a reduction factor $R$ as the ratio of the experimental cross section to theoretical prediction, usually $R<1$ due to correlations between the nucleons. Elaborated shell model (SM) calculations have not been able to explain the reductions obtained for $R$ or the quenching of spectroscopic factors (SF). An overall reduction of SFs compared to the SM has been observed, e.g., in Refs. \cite{TostevinPRC.90.057602,PhysRevC.103.054610}.

\subsection{Results and discussions}

In Tables II-XI we present our results for the calculations comparing them to the experimental data.

\begin{widetext}
\begin{center}
\begin{table}[!htb]
\centering
\caption{Proton(neutron) knockout with $^8$Be projectiles. }
\setlength{\tabcolsep}{2.96mm}{
\begin{tabular}{lcccccccc} \hline \hline
 Reaction          & $E_{beam}$  & $J^\pi$  &$j$& $S_n$&  $C^2S$(th)& $\sigma_{th}$&  $\sigma_{exp}$ & R \\  \hline
    & MeV/nucleon & && MeV&  &mb&mb\\  \hline
$^{12}$C($^{8}$B,$^{7}$Be) \cite{PhysRevC.67.064301} &76 & $3/2^-$ &3/2&0.137& 0.761 &53.98 & \\
 & & &$1/2$& & 0.073& 28.53\\
 & & $1/2^-$ &$3/2$&0.559& 0.227 &25.26 \\
 & & inclusive& &&&107.78& 130(11) & 1.206(102) \\
 $^{12}$C($^{8}$B,$^{7}$Be) \cite{BLANK1997242} &142 & $3/2^-$ &3/2&0.137& 0.761 &54.57 & \\
 & & &$1/2$& & 0.073& 19.87\\
 & & $1/2^-$ &$3/2$&0.559& 0.227 &21.75 \\
 & & inclusive& &&&96.19& 109(1) & 1.133(10) \\
 $^{12}$C($^{8}$B,$^{7}$Be) \cite{BLANK1997242} &285 & $3/2^-$ &3/2&0.137& 0.761 &47.64 & \\
 & & &$1/2$& & 0.073& 12.63\\
 & & $1/2^-$ &$3/2$&0.559& 0.225 &16.96 \\
 & & inclusive& &&&77.23& 89(2) & 1.154(13)\\
  $^{12}$C($^{8}$B,$^{7}$Be) \cite{2002PhLB52936C} &936 & $3/2^-$ &3/2&0.137& 0.761 &47.06 & \\
 & & &$1/2$& & 0.073& 15.26\\
 & & $1/2^-$ &$3/2$&0.559& 0.225 &17.64 \\
 & & inclusive& &&&79.96& 94(9) & 1.176(112) \\
   $^{12}$C($^{8}$B,$^{7}$Be) \cite{2001EPJA1049C}&1440 & $3/2^-$ &3/2&0.137& 0.761 &48.81 & \\
 & & &$1/2$& & 0.073& 18.07\\
 & & $1/2^-$ &$3/2$&0.559& 0.225 &19.19 \\
 & & inclusive& &&&86.07& 96(3) & 1.115 (34)\\
\hline \hline
\end{tabular}}\label{SF2}
\end{table}
\end{center}
\end{widetext}

In Table \ref{SF1} we show the comparison of our calculations and experimental data for proton(neutron) knockout from $^7$Li projectiles incident on $^9$Be targets. Our calculated spectroscopic factors and cross sections for the specified states are given in the sixth and seventh columns.  For the $^{9}$Be($^{7}$Li,$^{6}$He) reaction at 80 MeV/nucleon, our results are remarkably close to those reported in Ref. \cite{PhysRevC.86.024315} using wavefunctions calculated with the variational Monte Carlo (VMC) method \cite{PiperWiringa2001}. We get 28.13 mb, while the value quoted in Ref. \cite{PhysRevC.86.024315}  using VMC wavefunctions is is 26.7 mb. Our spectroscopic factor  for the $0^+$ state in the residual nucleus is 0.496 and 0.439 for the VMC method. The experimental  cross section is only 13.3(5) mb, one of the  smallest nucleon knockout reaction cross section probably related to  $^7$Li having a prominent $\alpha$-triton cluster structure in contrast to
$^6$He having a halo structure with an $\alpha$-core and two loosely-bound neutrons. The transition between the $^7$Li and $^6$He ground states through the knockout of a proton must be a fine-tuning reaction mechanism, not well described by  the reaction models neglecting couplings to other channels such as the dissociation of $^6$He as pointed out in Ref.   \cite{PhysRevC.86.024315}.

For the $^{9}$Be($^{7}$Li,$^{6}$Li) reaction at 120 MeV/nucleon our calculations yield 61.21 mb while with VMC wavefunctions one obtains 52.6 mb, and 53.8 with shell-model wavefunctions. The spectroscopic factors for the $1^+$ state are 0.649 compared to 0.715 with the VMC. For the $0^+$ state these numbers are 0.250 and 0.219, respectively. These values are much closer than those obtained with the shell-model calculations, in agreement with the analysis made in Ref. \cite{PhysRevC.86.024315}. The small differences found shouldn't be ascribed to the reaction calculations but to the details of the ab initio wavefunctions, which do not perfectly agree with each other. But it is worth noticing that the calculations using shell-model, VMC and NCSM wavefunctions overestimate the cross sections compared to the experimental value of 30.7 mb. There is no clear explanation for the disagreement between theory and experiment.

In Table \ref{SF2} we show the comparison of our calculations and experimental data for proton knockout from $^8$B projectiles incident on $^{12}$C targets at different beam energies. Our calculated spectroscopic factors and cross sections for the specified states are given in the sixth and seventh columns. The proton is assumed to be removed from either the $j= 3/2$ or the $j=1/2$ orbital of the $^8$B ground state. In our model, the $2^+$ ground state in $^8$B is an admixture of contributions from the 3/2 and 1/2 orbitals. The  residual nucleus $^7$Be is left either in its  $3/2^-$ ground state or in its $1/2^-$ excited  state at 429 keV.  It is worthwhile comparing our results with the theoretical calculations for the same reactions (except the one for 76 MeV/nucleon) as reported in Ref. \cite{BrownPRC.65.061601}. We notice the same bombarding energy dependence of the cross sections, decreasing to a minimum at 285 MeV/nucleon before increasing again. This  is an expected feature reminiscent of the energy dependence of the  nucleon-nucleon cross section. But our cross sections are consistently smaller, by about 20\% than the theoretical results obtained with shell-model wavefunctions reported in Ref. \cite{BrownPRC.65.061601}.

One of the possible explanations for this disagreement is the very small proton separation energy. It is often stated that the NCSM wavefunctions do not properly describe the long tails of loosely-bound states in light nuclei because they are based on an expansion in harmonic oscillator basis which naturally display a quick fall off at large distances. However, the internucleon correlations are well treated within the NCSM. This was indeed shown in Refs. \cite{Navratil:06plb,Navratil:06prc} where the problem was fixed by matching the NCSM wavefunctions with the expected tails  well described by Whittaker functions. This trick allowed a good reproduction of the momentum distributions and cross sections for knockout reactions with $^8$B projectiles.  Another noteworthy observation is that our spectroscopic factors are very close to those shell-model results in Ref. \cite{BrownPRC.65.061601}, except for that of the $j=3/2$ and the $3/2^-$ ground state of $^7$Be. Our numerical value is 0.761 while Ref. \cite{BrownPRC.65.061601} indicated a spectroscopic factor of 0.97. The model used in Ref. \cite{BrownPRC.65.061601} is directly related to the spectroscopic factor (see Eq. (1) in Ref. \cite{BrownPRC.65.061601}). In the present work, the  Glauber model \cite{Hansen:03,Bertulani:06} is used, in which the overlap integrals acts as the input information. This could be a possible reason for the difference.

\begin{widetext}
\begin{center}
\begin{table}[!htb]
\centering
\caption{Knockout reactions with $A =9$ projectiles}
\setlength{\tabcolsep}{2.96mm}{
\begin{tabular}{lcccccccc} \hline \hline
 Reaction          & $E_{beam}$  & $J^\pi$  &$j$& $S_p$[$S_n$]&  $C^2S$(th)& $\sigma_{th}$&  $\sigma_{exp}$ & R \\  \hline
    & MeV/nucleon & && MeV&  &mb&mb\\  \hline
$^{12}$C($^{9}$C,$^{8}$B) \cite{PhysRevC.67.064301} & 78 & $2^+$ &3/2& 1.3&0.876&42.24 & \\
 &  &  &1/2& &0.042& 13.96& \\
  &  & inclusive && &&56.2 &54 (4) & 0.961(71)\\
$^{9}$Be($^{9}$C,$^{8}$B) \cite{PhysRevC.86.024315} & 100 & $2^+$ &3/2& 1.3&0.876&46.21 & \\
 &  &  &1/2& &0.042& 10.68& \\
  &  &inclusive  &&  &&56.9 &56 (3) & 0.984(53) \\
 $^{9}$Be($^{9}$Li,$^{8}$Li) \cite{PhysRevC.86.024315} & 80 & $2^+$ &3/2& 4.06&0.895&36.38 & \\
 &  &  &1/2& &0.042& 8.758& \\
&  & $1^+$ &3/2& 5.05&0.026&5.423 & \\
  &  &inclusive  && &&50.56 &55.6(29) & 1.100(57) \\
   $^{12}$C($^{9}$Li,$^{8}$Li) \cite{PhysRevC.86.024315} & 100 & $2^+$ &3/2& 4.06&0.895&47.82 & \\
 &  &  &1/2& &0.040& 10.94& \\
&  & $1^+$ &3/2& 5.05&0.026&6.90 & \\
  &  & inclusive && &&65.66 &62.9(41) & 0.958(62) \\
\hline \hline
\end{tabular}}\label{SF3}
\end{table}
\end{center}
\end{widetext}

In Table \ref{SF3} we show the comparison of our calculations and experimental data for proton knockout from $^9$C projectiles incident on $^{12}$C targets. Our calculated spectroscopic factors and cross sections for the specified states are given in the sixth and seventh columns. The proton is assumed to be removed from either the $j= 3/2$ or the $j=1/2$ orbital of the $^9$C $3/2^-$ ground state with a proton separation energy of 1.3 MeV. Both reactions $^{12}$C($^{9}$C,$^{8}$B)X and $^{9}$Be($^{9}$C,$^{8}$B)X have similar experimental values and our calculations using NCSM wavefunctions are in excellent agreement with the experiments. According to Ref. \cite{PhysRevC.86.024315} the proton knockout $^{9}$Be($^{9}$C,$^{8}$B)X at 100 MeV/nucleon calculated with VMC wavefunctions \cite{PiperWiringa2001}  is 64.4(15) mb, about 20\% larger than the experimental values.  The cross sections calculated with the shell-model for $^{12}$C($^{9}$C,$^{8}$B)X at 78 MeV/nucleon is 65.7 mb, as reported in Ref. \cite{PhysRevC.67.064301}. Our results are much closer to the experimental data, yielding a much smaller reduction factor $R=\sigma_{exp}/\sigma_{th}$.

The reaction ($^9$Li,$^8$Li) is mirror symmetric with respect to ($^9$C,$^8$B) and the measured and calculated cross sections are nearly equal, despite an additional excited state in $^8$Li. Because of the mirror symmetry, the spectroscopic factors for the $2^+$ state are almost identical. But one notices that the neutron removal from the $1^+$ state adds an extra 6 mb to the ($^9$Li,$^8$Li) reaction, which is absent in the ($^9$C,$^8$B) reaction. This is substantial smaller than the 20 mb value for the removal form the $1^+$ state reported in Ref. \cite{PhysRevC.86.024315}. Therefore, our calculations are in accordance with the mirror symmetry and the expected cross sections are concentrated in the same $2^+$ state, as expected from the symmetry.

\begin{widetext}
\begin{center}
\begin{table}[!htb]
\centering
\caption{Knockout reactions with $A =10$ projectiles.}
\setlength{\tabcolsep}{1.26mm}{
\begin{tabular}{lcccccccccc} \hline \hline
 Reaction          & $E_{beam}$  & $J^\pi$  &$j$& $S_p$[$S_n$]&  $C^2S$(th)& $C^2S$(VMC)&  $C^2S$(SM) & $\sigma_{th}$&  $\sigma_{exp}$ & R \\  \hline
    & MeV/nucleon & && MeV& && &mb&mb\\  \hline
$^{9}$Be($^{10}$Be,$^{9}$Li) \cite{PhysRevC.86.024315} & 80 & $3/2^-$ & $3/2$ &19.64&1.375 & 1.043 &1.929  &41.37& \\
 &  & $1/2^-$ & $1/2$ &22.33& 0.424& 0.434 &0.282 & 18.30& \\
 &  & inclusive & &&&& &59.67&26.0(13) & 0.435(22)\\
$^{9}$Be($^{10}$Be,$^{9}$Be) \cite{PhysRevC.86.024315} & 120 & $3/2^-$ & 3/2 & 6.812& 2.148& 1.963 & 2.622& 88.08& 71.2(40)  & 0.810(48) \\
& 80 & &   & & && & 66.54& 69.5(32) & 1.045(45)\\
$^{9}$Be($^{10}$C,$^{9}$C) \cite{PhysRevC.86.024315} & 120 & $3/2^-$ &  3/2 & 21.28&  1.340& 1.043 & 1.933 & 48.98 &23.4(11)  & 0.478(22) \\
$^{12}$C($^{10}$C,$^{9}$C) \cite{PhysRevC.86.024315} & 120 & $3/2^-$ &  3/2 & 21.28 & 1.340 & 1.043 & 1.933 &55.63 &27.4(13) & 0.492(23)\\
\hline \hline
\end{tabular}}\label{SF4}
\end{table}
\end{center}
\end{widetext}

In Table \ref{SF4} we show the comparison of our calculations and experimental data for knockout reactions with $A =10$ projectiles on $^9$Be and $^{12}$C targets. The ($^{10}$Be,$^9$Li) proton knockout reaction, the mirror reaction to ($^{10}$C,$^9$C). The ($^{10}$Be,$^9$Li) reaction with a proton removal from the $0^+$ state can populate the $3/2^-$ ground state of $^9$Li, as well as its excited state at 2.69 MeV. Our spectroscopic factors are similar to the VMC calculations reported in Ref. \cite{PhysRevC.86.024315}, although for the $3/2^-$ states they are about 20\% smaller, whereas the VMC and shell-model spectroscopic factors differ by up to a factor of 2. However, our calculated cross section is about the same percentage larger than the VMC.  The VMC theoretical cross section adds up to 50.3 mb, whereas our result is 59.67 mb. Both calculations over-predict the experimental value of 26.0(13) mb, as observed in Ref. \cite{PhysRevC.86.024315}.

The $(^{10}$Be,$^9$Be) reaction with a neutron removal from the $0^+$ state can populate the $3/2^-$ ground state of $^9$Be, with a neutron separation energy of 6.81 MeV.  Our spectroscopic factor of 2.148 lies in-between the shell-model and the VMC respective values of 2.622  and 1.932 reported in Ref. \cite{PhysRevC.86.024315}. The cross sections obtained with our wavefunction are 66.54 mb and 88.08 mb, at 80 MeV/nucleon and 120 MeV/nucleon, respectively. These results also lie in-between the shell-model and VMC values and are in reasonable agreement with the experimental data.

The $(^{10}$C,$^9$C) reaction with a neutron removal from the $0^+$ state can populate the $3/2^-$ ground state of $^9$C, with a neutron separation energy of 21.28 MeV.  Our spectroscopic factor of 1.340 once more lies inbetween the shell-model and the VMC respective values of 1.933  and 1.043 reported in Ref. \cite{PhysRevC.86.024315}. The cross sections obtained with our wavefunction is 48.98 mb for Be targets and 120 MeV/nucleon $^{10}$C projectiles and 55.63 mb and for C targets. These results are closer to those obtained with VMC wave functions, but are again larger by about a factor of 2 than the experimental data.

\begin{widetext}
\begin{center}
\begin{table}[!htb]
\centering
\caption{Knockout reactions with $A =12$ projectiles}
\setlength{\tabcolsep}{2.96mm}{
\begin{tabular}{lcccccccc} \hline \hline
 Reaction          & $E_{beam}$  & $J^\pi$  &$j$& $S_p$$[S_n]$&  $C^2S$(th)& $\sigma_{th}$&  $\sigma_{exp}$ & R \\  \hline
    & MeV/nucleon & && MeV&  &mb&mb\\  \hline
$^{12}$C($^{12}$C,$^{11}$B) \cite{PhysRevC.37.2613} &250 & $3/2^-$&3/2 &15.95& 2.690& 77.31 &65.6(26) & 0.849(34) \\
$^{12}$C($^{12}$C,$^{11}$B) \cite{PANIN2019134802} &400 & &&& & 67.44 &60.9(27) & 0.903(40) \\
 $^{12}$C($^{12}$C,$^{11}$B) \cite{PhysRevC.28.1602}&1050 & &&& & 61.89&48.6(24) & 0.785(38)\\
&2100 & &&& & 62.58&53.8(27)& 0.860(43) \\
 $^{12}$C($^{12}$C,$^{11}$C) \cite{PhysRevC.37.2613} &250 & $3/2^-$&3/2 &18.72& 2.497& 78.19 &56.0(41) & 0.716(52)  \\
  $^{12}$C($^{12}$C,$^{11}$C)  \cite{PhysRevC.28.1602}&1050 & &&& & 61.54&44.7(28) & 0.726(45) \\
 &2100 & &&& & 62.03&46.5(23) & 0.750(37)\\
\hline \hline
\end{tabular}}\label{SF5}
\end{table}
\end{center}
\end{widetext}

In Table \ref{SF5} we compare our calculations to the inclusive reaction cross sections for nucleon removal from $^{12}$C projectiles incident on carbon targets at 250, 400, 1050 and 2100 MeV/nucleon, as reported in Refs. \cite{PhysRevC.37.2613,PANIN2019134802,PhysRevC.28.1602}.  All particle removals are assumed to take place from the $p_{3/2}$ orbitals. But it is worthwhile mentioning that all these data are inclusive, with no identification of the orbitals from which the nucleons are removed. Our calculated spectroscopic factors and cross sections for the specified states are given in the sixth and seventh columns.  It is noticeable that the calculated cross sections are about 10-20\% larger for the proton removal case of $^{12}$C($^{12}$C,$^{11}$B) and substantially larger for neutron removal cases of $^{12}$C($^{12}$C,$^{11}$C). Nonetheless, our theoretical calculations yield very similar results as those reported in previous references, as in Ref. \cite{PhysRevC.92.024614} where a quantum molecular dynamics (QMD) model was used for comparison with the experimental data. The physics included in our theoretical model is not easy to translate into those included in QMD. The QMD model assumes wave packets for nucleons incorporating momentum-dependent forces, surface tension, and a Pauli ``force" into the Hamiltonian as effective potential terms. The model is popular in investigate multi-fragmentation reactions in nucleus-nucleus collisions. A common problem with the model is the inaccurate treatment of peripheral collisions attributed to spurious excitation and disintegration of nuclei. Such effects could not be distinguished from the true events arising from peripheral collisions \cite{PhysRevC.41.547}. These drawbacks were apparently fixed in Ref.  \cite{PhysRevC.92.024614} and it is remarkable that despite the very different modeling, their results are in the same ballpark as ours.
\begin{widetext}
\begin{center}
\begin{table}[!htb]
\centering
\caption{Knockout reactions with $A =14$ projectiles}
\setlength{\tabcolsep}{2.56mm}{
\begin{tabular}{lccccccccc} \hline \hline
 Reaction          & $E_{beam}$  & $J^\pi$  &$j$& $S_p$$[S_n]$&  $C^2S$(th)&  $C^2S$(SM) & $\sigma_{th}$&  $\sigma_{exp}$ & R \\  \hline
    & MeV/nucleon & && MeV& & &mb&mb\\  \hline
    $^{12}$C($^{14}$O,$^{13}$N) \cite{PhysRevC.90.037601} & 305   &$1/2^-$ & $1/2$&4.63 &1.531 & 1.55 &39.66  &  35(5) & 0.883(126) \\
     $^9$Be($^{14}$O,$^{13}$N) \cite{PhysRevLett.108.252501} & 53   & &&& &&36.29  &  58(4) & 1.607(110) \\
          $^9$Be($^{14}$O,$^{13}$O) \cite{PhysRevLett.108.252501} & 53   &$3/2^-$ &$3/2$&23.18 & 3.234 & 3.15 & 31.34  &  14(4) & 0.465(127) \\
\hline \hline
\end{tabular}}\label{SF6}
\end{table}
\end{center}
\end{widetext}

In Table \ref{SF6} we compare our calculations to the  inclusive cross sections for nucleon removal from $^{14}$O projectiles incident on beryllium and carbon targets at 53 and 305 MeV/nucleon, as reported in Refs. \cite{PhysRevC.90.037601,PhysRevLett.108.252501}. We assume that the neutrons are removed from the $p_{3/2}$  orbital, whereas the proton is removed from the $p_{1/2}$ orbital. Our calculated spectroscopic factors and cross sections for the specified states are given in the sixth and seventh columns. Calculations based on the intra-nuclear cascade method \cite{PhysRevC.83.011601} predict for the $^{12}$C($^{14}$O,$^{13}$N) reaction at 305 MeV/nucleon a cross section of  39 MeV, assuming a scaled geometry, as explained in Ref. \cite{PhysRevC.90.037601}. This is amazingly close to our calculated value of 39.66 mb, despite the differences in the reaction models used. However, the experimental cross section, as reported in Ref. \cite{PhysRevLett.108.252501} for the $^9$Be($^{14}$O,$^{13}$N) reaction at 53 MeV/nucleon is 58(4) mb, contrasting sharply with the one published Ref. \cite{PhysRevC.90.037601}.  The cross section for $^{12}$C($^{14}$O,$^{13}$N) at 53 MeV/nucleon as reported in Ref. \cite{PhysRevLett.108.252501} is very different from our calculated value and also from their own eikonal calculations. Our calculated value of 36.29 mb for this reaction is consistent with the expectation that is the energy dependence of the nucleon-nucleon cross section, which decreases rapidly as the beam energy increases from 50 MeV/nucleon to 300 MeV/nucleon. Since the eikonal model used here yields similar cross sections as the intra-nuclear cascade model at 305 MeV/nucleon, it is hard to pinpoint where the source of discrepancy might be. For more discussions on that, see Ref. \cite{PhysRevC.90.037601}.
The situation is inverted for the  $^9$Be($^{14}$O,$^{13}$O)  reaction at 53 MeV/nucleon, with our calculated cross section value of 31.34 mb whereas the experimental cross section is quoted being 14(4) mb.
Both our calculations of the spectroscopic factor and cross-section are close to the results in Ref. \cite{PhysRevLett.108.252501,Flavigny:2012}.

\begin{widetext}
\begin{center}
\begin{table}[!htb]
\centering
\caption{\label{RE} Knockout reactions with $A =16$ projectiles}
\setlength{\tabcolsep}{2.96mm}{
\begin{tabular}{lcccccccc} \hline \hline
 Reaction          & $E_{beam}$  & $J^\pi$  &$j$& $S_p$($S_n$)&  $C^2S$(th)& $\sigma_{th}$&  $\sigma_{exp}$ & R \\  \hline
    & MeV/nucleon & && MeV&  &mb&mb\\  \hline
$^{12}$C($^{16}$O,$^{15}$N) \cite{PhysRevC.37.2613} & 2100   &  $1/2^-$&$ 1/2$  &12.13&1.581 &28.83&   \\
 &   &  $3/2^-$&  $3/2$&22.04& 3.066 &49.84&   \\
  &   & inclusive &  &&&78.67&  54.2(29)  & 0.689(37) \\
$^{12}$C($^{16}$O,$^{15}$O) \cite{PhysRevC.28.1602} & 2100  &  $1/2^-$&$ 1/2$  &12.13&1.702&30.41&    \\
&  &  $3/2^-$&$ 3/2$  &22.04& 3.209&47.58&    \\
 &   &inclusive  &  &&&77.99&  42.9(23) & 0.550(29) \\
\hline \hline
\end{tabular}}\label{SF7}
\end{table}
\end{center}
\end{widetext}

In Table \ref{SF7} we compare our calculations to the inclusive cross sections for nucleon removal from $^{16}$O projectiles incident on beryllium and carbon targets at 53 and 305 MeV/nucleon, as reported in Refs. \cite{PhysRevC.37.2613,PhysRevC.28.1602}. The nucleons are removed from the $p_{3/2}$ and $p_{1/2}$ orbitals. Our calculated spectroscopic factors and cross sections for the specified states are given in the sixth and seventh columns. The calculated values are in rather good agreement with the experimental data for the reaction $^{12}$C($^{16}$O,$^{15}$N) at 2100 MeV/nucleon, whereas an appreciably larger enhancement factor of the theoretical cross section is observed for the neutron removal reaction $^{12}$C($^{16}$O,$^{15}$O) at the same energy.

\begin{widetext}
\begin{center}
\begin{table}[!htb]
\centering
\caption{\label{RE} Quasi-free (p,2p) and (p,pn) reactions with $^{12}$C.}
\setlength{\tabcolsep}{2.96mm}{
\begin{tabular}{lcccccccc} \hline \hline
 Reaction          & $E_{beam}$  & $J^\pi$  &$j$& $S_p(S_n)$&  $C^2S$(th)& $\sigma_{th}$&  $\sigma_{exp}$ & R\\  \hline
    & MeV/nucleon & && MeV&  &mb&mb\\  \hline
$^{12}$C(p,2p)$^{11}$B \cite{Pan16} & 400   &  $3/2^-$&$ 3/2$  &15.96&2.497&16.23&15.8(18)   \\
 &   &  $1/2^-$&  $1/2$&18.08&0.475&2.492&1.9(2)   \\
 &   &  inclusive&  $$&&&18.72&17.7(18)  & 0.945(96)  \\
 $^{12}$C(p,pn)$^{11}$C \cite{HOLL2019682}& 400   &  $3/2^-$&$ 3/2$  &18.72&2.686&24.16&\\
&   &  $1/2^-$&$ 1/2$  &21.40&0.512&4.457&\\
 &   &  inclusive&  $$&&& 28.62&30.0(32)(27) & 1.048(206)  \\
\hline \hline
\end{tabular}}\label{SF8}
\end{table}
\end{center}
\end{widetext}

We now turn on to quasi-free reactions of the (p,2p) and (p,pn) type in inverse kinematics. In Table \ref{SF8} we compare our calculations to the quasi-free cross sections for $^{12}$C(p,2p)$^{11}$Be and  $^{12}$C(p,pn)$^{11}$C reaction reported in Refs. \cite{Pan16,HOLL2019682}. The reaction part of the calculations follow the theory developed in Ref. \cite{Aumann:13}. The nucleons are assumed to be removed from the $1p_{3/2}$ and $1p_{1/2}$ orbitals, for both cases. For the $^{12}$C(p,2p)$^{11}$Be reaction the sum of our calculated spectroscopic factors is equal to 2.97 whereas the value quoted in Ref.  \cite{HOLL2019682} is $\sum C^2S = 4.28$ based on the shell-model calculated with  the Warburton-Brown interaction \cite{PhysRevC.46.923} in the $spsdpf$ model space restricted to $(0+1)\hbar \omega$. The agreement with our results are outstanding. But the summed set of spectroscopic factors are very different. We notice that in Refs. \cite{Pan16,HOLL2019682} the calculations reported were done using a Woods-Saxon wavefunction and spectroscopic factors from shell-models. In our case, we have used wavefunctions calculated with the no-core-shell model for consistency. As recently discussed in Ref. \cite{BertulaniIdini}, the quasi-free cross sections are strongly dependent of the form of the wavefunctions, and not only the spectroscopic factors. Therefore, we deem the agreement between the two sets of calculations (shell model and ab initio) as being coincidental for these two reactions.

\begin{widetext}
\begin{center}
\begin{table}[!htb]
\centering
\caption{ (p,2p) reactions with oxygen isotopes. Experimental cross sections are from Ref. \cite{Atar:2018}, listed along with statistical (round brackets) and systematic uncertainties (square brackets). }
\begin{tabular}{lcccccccc} \hline \hline
 Reaction          & $E_{beam}$  & $J^\pi$  &$j$& $S_p$&  $C^2S$(th)& $\sigma_{th}$&  $\sigma_{exp}$ & R\\  \hline
    & MeV/nucleon & && MeV&  &mb&mb\\  \hline
$^{14}$O(p,2p)$^{13}$N  \cite{Atar:2018}& 351  &  $1/2^-$ &$1/2$&4.626& 1.532& 16.58& 10.23(0.80)[0.65]  & 0.617(87)\\
$^{16}$O(p,2p)$^{15}$N   \cite{Atar:2018}& 451  & $1/2^-$ &1/2  &12.13&1.581& 12.24&   \\
  &  &   $3/2^-$ &3/2  &18.76& 3.209 &16.69 &    \\
&&inclusive & &  &  &28.93&    26.84(0.90)[1.70] & 0.928(89) \\
\hline \hline
\end{tabular}\label{SF9}
\end{table}
\end{center}
\end{widetext}

To reinforce the arguments raised above, in Table \ref{SF9} we list cross sections for (p,2p) reactions with oxygen isotopes. Experimental cross sections are from Ref. \cite{Atar:2018}, presented along with statistical (round brackets) and systematic uncertainties (square brackets). All protons are assumed to be removed from the $1p_{1/2}$ orbital, except for the $^{16}$O case, where they are assumed to be removed from a mixture of the $1p_{1/2}$ orbital (ground state) and from $1p_{3/2}$ orbital corresponding to two excited states at 6.63 and 9.93 MeV, with the inclusive cross section listed for the sum of their contributions. For $^{16}$O(p,2p)$^{15}$N, the states with separation energies 18.76 MeV and 22.06 MeV are considered as part of a single mixed state. As in the previous data of table \ref{SF8}, the calculations reported in Ref. \cite{Atar:2018} do not use ab initio wavefunctions but those generated with Woods-Saxon potentials scaled with spectroscopic factors for the ab initio self-consistent Green's function (SCGF) method. Our calculated cross sections use not only the spectroscopic factors from the ab initio NCSM but also the wavefunctions generated in the model. Our results  are about 20\% smaller than the theoretical results using SCGF spectroscopic factors \cite{doi:10.1142/S0217751X09045625,Cipollone:15} as reported in Ref. \cite{Atar:2018}. This again reinforces the idea of the importance of using proper ab initio wavefunctions for a consistent analysis of the experimental data.

\begin{center}
\begin{table}[!htb]
\centering
\caption{The computed reduction factor, R, $\Delta S=S_p-S_n$ for neutron removal or $\Delta S=S_p-S_n$ for proton removal.}
\setlength{\tabcolsep}{1.8mm}{
\begin{tabular}{lccc} \hline \hline
 Reaction           & $\Delta  S$ & $E_{beam}$ & R \\  \hline
                   & MeV & MeV/nucleon &    \\ \hline
$^{9}$B($^7$Li,$^6$He) & 2.723  & 80  &  0.476(24)  \\
$^{9}$B($^7$Li,$^6$Li)  & -2.723 & 120  &  0.501(29)  \\
$^{12}$C($^8$B,$^7$Be)  &  -12.690  & 76   &  1.206(102)  \\
                        & & 142             & 1.133(10)  \\
                        & & 285            & 1.154(13)  \\
                        & & 936              &1.176(112)  \\
                        & & 1440             & 1.115(34) \\
$^{12}$C($^9$C,$^8$B)  & -12.925  & 78   &  0.961(71)  \\
$^{9}$Be($^9$C,$^8$B)  &  -12.690  & 100   &  0.984(53)  \\
$^{9}$Be($^9$Li,$^8$Li)  &  -9.882  & 80   &  1.100(57)  \\
$^{12}$C($^9$Li,$^8$Li)  &  -9.882  & 100   &  0.958(62)  \\
$^{9}$Be($^{10}$Be,$^9$Li)  &  12.824  & 80   &  0.435(22)  \\
$^{9}$Be($^{10}$Be,$^9$Be)  &  -12.824  & 80   &  1.045(45)  \\
                            &            & 120  & 0.810(48) \\
$^{9}$Be($^{10}$C,$^9$C)  &  17.277  & 120   &  0.478(22)  \\
$^{12}$C($^{10}$C,$^9$C)  &  17.277  & 120   &  0.492(23)  \\
$^{12}$C($^{12}$C,$^{11}$B)  & -2.764  & 250   &  0.849(34)  \\
                             &         & 400   &0.903(40) \\
                             &         & 1050   &0.785(38) \\
                             &         & 2100   &0.860(43) \\
$^{12}$C($^{12}$C,$^{11}$C)  &  2.764  & 250   &  0.716(52)  \\
                             &         & 1050   &0.726(45) \\
                             &         & 2100   &0.750(37) \\
$^{12}$C($^{14}$O,$^{13}$N)  &  -18.552  & 305   &  0.883(126)  \\
$^{9}$Be($^{14}$O,$^{13}$N)  &  -18.552  & 53   &  1.607(110)  \\
$^{9}$Be($^{14}$O,$^{13}$O)  &   18.552  & 53   &  0.465(127)  \\
$^{12}$C($^{16}$O,$^{15}$N)  &   -3.537  & 2100   &  0.689(37)  \\
$^{12}$C($^{16}$O,$^{15}$O)  &    3.537  & 2100   &  0.550(29)  \\
$^{12}$C(p,2p)$^{11}$B  &    -2.764  & 400   &  0.945(96)  \\
$^{12}$C(p,pn)$^{11}$C  &     2.764  & 400   &  1.048(206)  \\
$^{14}$O(p,2p)$^{13}$N  &    -18.552  & 351   &  0.617(87)  \\
$^{16}$O(p,2p)$^{15}$N  &    -3.537  & 451   &  0.928(89)  \\
\hline \hline
\end{tabular}}\label{SF10}
\end{table}
\end{center}

\begin{figure}[!htb]
\includegraphics[width=1.00\columnwidth]{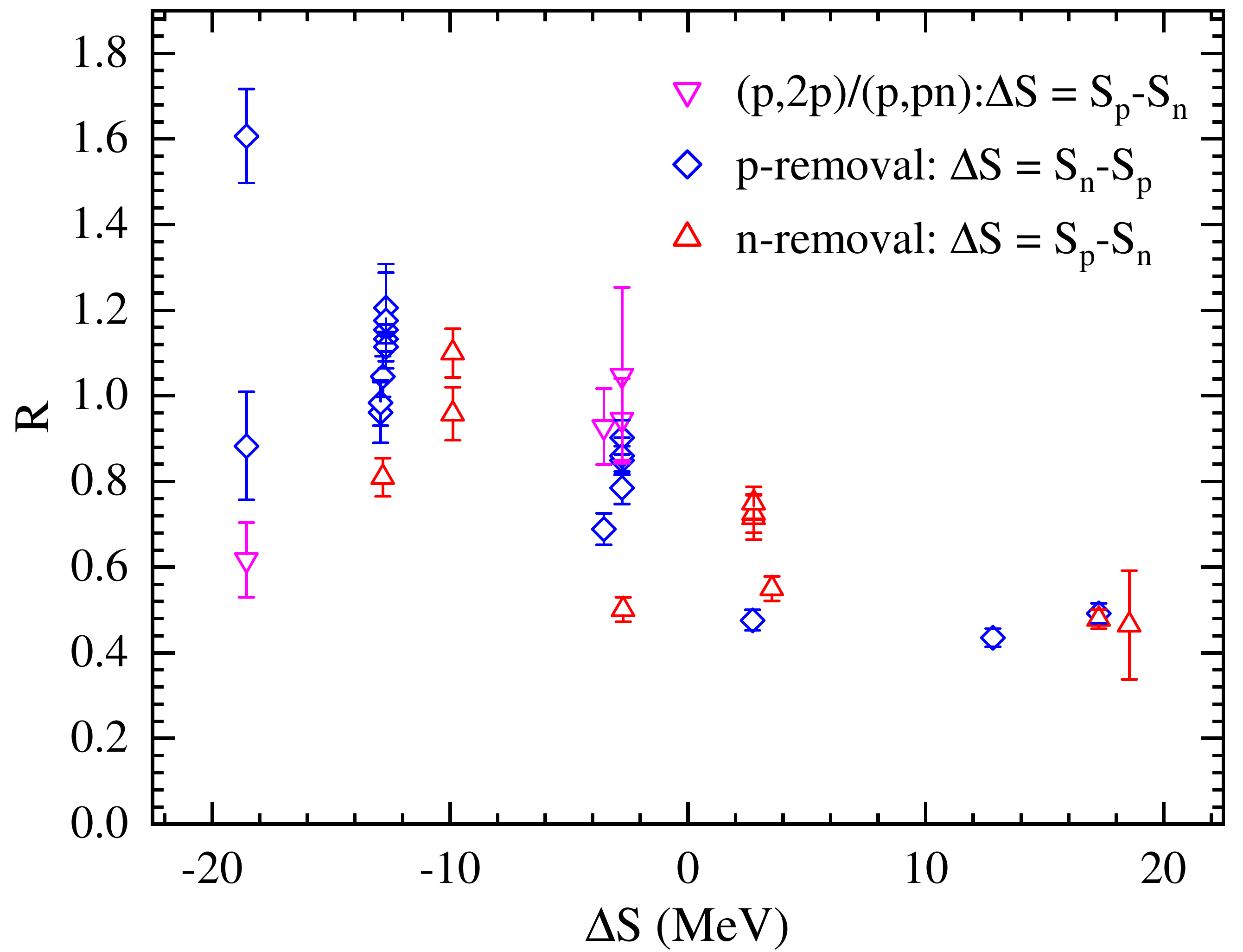}
\caption{Compilation of the computed reduction factors between the experimental and  theoretical  inclusive one-nucleon-removal cross sections. Blue, red and pink symbols are for the proton removal, neutron removal and (p,2p) [or (p, pn)], respectively. }{\label{fig}}
\end{figure}

\section{Conclusions}
$(p,pN)$ probes are much more sensitive to the details of the internal part of the overlap functions than knockout by heavy targets.  But the later type of reactions is also influenced by the internal details of the overlap functions.  A close inspection of Tables \ref{SF1}-\ref{SF9}  reveals that substantial agreements exist in the experimental reduction factors for heavy ion induced reactions obtained with wavefunctions generated with  many-body overlap functions.
The calculated reduction factors for reactions investigated in the present work are summarized in Fig. \ref{fig}  and Table \ref{SF10}.
There is a small tendency of the reduction factor for (p,pN) reactions to increase with  $\Delta S$, whereas this trend is reversed for heavy ion knockout reactions, in accordance with the findings of Refs.  \cite{PhysRevC.103.054610} and \cite{PhysRevC.90.057602}.
Further, we have been able to reinforce the notion that nucleon removal in heavy-ion knockout reactions cannot be only ascribed to the asymptotic behavior of the wavefunctions. A simple rescaling of the tails of the wavefunction with an ANC or by multiplication with spectroscopic factors can lead to a misidentification of important nuclear structure effects imbedded in the overlap functions. Therefore, the whole picture of quenching of spectroscopic factors  might be misleading if Woods-Saxon wavefunctions, or wavefunction tails, are used in the analysis. This assertion indicates that future experimental analyses of nucleon removal in (p,pN) and heavy ion knockout reactions require a closer collaboration of experiment and ab initio theory practitioners than typically reported in the literature.

In summary, we have shown that a proper experimental analysis  requires the input of a properly calculated overlap function from first principles. While this poses a more difficult task for the study of single-particle configurations with $(p,pN)$ reactions, it also opens opportunities for a better understanding of the nucleon-nucleon correlation effects in the overlap functions.

\acknowledgments{
The NCSM code was provided by N. Michel, which is publicly available on \url{https://github.com/GSMUTNSR}. We also acknowledge J. P. Vary and P. Maris for providing us the
Daejeon16 interaction.
This work has been supported by the National Key R\&D Program of China under Grant No. 2018YFA0404401; the National Natural Science Foundation of China under Grants No. 11835001,  11921006, 12035001, and  11975282; the State Key Laboratory of Nuclear Physics and Technology, Peking University under Grant No. NPT2020KFY13; the Strategic Priority Research Program of Chinese Academy of Sciences under Grant No. XDB34000000; and the CUSTIPEN (China-U.S. Theory Institute for Physics with Exotic Nuclei) funded by the U.S. Department of Energy, Office of Science under Grant No. de-sc0009971; the U.S.\ DOE grant DE- FG02-08ER41533. We acknowledge the High-Performance Computing Platform of Peking University for providing computational resources.}

\bibliography{nuclear}



\end{document}